\documentclass[]{article}
\usepackage{amsmath}
\numberwithin{equation}{section}
\usepackage{amssymb}
\usepackage{graphicx}
\usepackage{authblk}
\usepackage[authoryear,round]{natbib}
\usepackage{hyperref}
\usepackage{relsize}
\usepackage{mathtools, nccmath}
\usepackage{xcolor}
\usepackage{cancel}
\usepackage{float}
\usepackage{array, booktabs, makecell}
\usepackage{siunitx}
\usepackage{physics}
\usepackage{placeins}
\usepackage{graphicx}
\graphicspath{ {./Comparison Plot/}}
\title{Spatial Phonons: A Phenomenological Viscous Dark Energy Model for DESI}

\author[1,2]{Muhammad Ghulam Khuwajah Khan\thanks{\href{mailto:b24bs1234@iitj.ac.in}{b24bs1234@iitj.ac.in}}\thanks{\href{khanmuhammadghulam@rjcollege.edu.in}{khanmuhammadghulam@rjcollege.edu.in}}}

\affil[1]{\textit{Department of Physics, Ramniranjan Jhunjhunwala College, Mumbai 400\,086, Maharashtra, India} \vspace{0.8em}}

\affil[2]{\textit{School of Artificial Intelligence and Data Science, Indian Institute of Technology Jodhpur, Jodhpur 342\,037, Rajasthan, India}}

\date{}
\begin{document}

\maketitle

\begin{abstract}
	We explore a phenomenological model of dark energy in which space is treated as an elastic brane with uniform tension $T_s$ and supports a longitudinal phonon sector described by three scalar fields $\phi^I$. At the background level the construction reproduces a perfect fluid whose enthalpy and bulk modulus are controlled by two dimensionless parameters $\varepsilon$ and $\kappa$, which encode the elastic and viscous response and determine both the effective equation of state $w_{\rm eff}(z)$ and the physical sound speed thriugh $c_s^2 = \kappa/\varepsilon$. Motivated by the consistency requirements of a phonon effective description, we focus on the stable regime $0 < \kappa < \varepsilon < 1$, which prevents tachyonic behavior and is natural given that $\varepsilon$ and $\kappa$ are defined as fractional response coefficients. To connect the model predictions to current observational summaries, we map $w_{\rm eff}(z)$ onto the Chevallier--Polarski--Linder form $w(z) = w_0 + w_a z/(1+z)$ over the redshift range relevant to DESI. We then use the public DESI DR1 compressed Gaussian likelihood in $(w_0,w_a)$ for BAO combined with Pantheon+ Type I a supernovae and Planck 2018, and scan the model parameter space to identify viable regions. We find that the model can reproduce the DESI compressed constraints at the background level, with representative best match values $(w_0,w_a) \simeq (-0.83,-0.74)$ and therefore, a near luminal sound speed $c_s^2 \simeq 0.9618$ as an output of the fit rather than a built in restriction. The covariance weighted distance is found to satisfy $\chi^2 \ll 1$ in the $(w_0,w_a)$ plane. The associated phonon mass scale $m_\phi = H_\star/(2\pi)$ is ultralight, corresponding to a Compton wavelength of order the cosmological horizon, consistent with interpreting the relevant excitations as infrared collective modes of the brane.
\end{abstract}

\newpage
\section{\centering Introduction}

Over the last quarter century a coherent observational picture has emerged in which the expansion of the Universe is accelerating. The first direct evidence came from type Ia supernovae used as standardizable candles at redshifts of order unity, which appeared dimmer than expected in a decelerating Universe (\cite{Riess1998, Perlmutter1999}). Subsequent measurements of the cosmic microwave background anisotropies and of baryon acoustic oscillations established a spatially flat cosmology in which a dominant smooth component with negative effective pressure drives the late time acceleration (\cite{Eisenstein2005BAO, Copeland2006DEReview, CaldwellKamionkowski2009, Planck2018}). 
\\ \\
In most observational analyses this component is described phenomenologically by an equation of state parameter $w(z) \equiv p_{\rm DE}/\rho_{\rm DE}$, with $\Lambda$CDM corresponding to $w = - 1$. Mild departures are commonly captured by low dimensional parametrizations such as the Chevallier--Polarski--Linder form (\cite{ChevallierPolarski2001,Linder2003}),
\begin{equation}
	w(z) = w_0 + w_a \, \frac{z}{1+z} \, ,
\end{equation}
which is flexible enough to represent a wide range of smooth evolutions while remaining directly constrainable by current datasets. 
\\ \\
The Dark Energy Spectroscopic Instrument has recently delivered the first year cosmological constraints from DESI baryon acoustic oscillation measurements across $z \simeq 0.1$ to $z \simeq 4$ (\cite{DESI2024VI}). BAO-only constraints remain statistically consistent with flat $\Lambda$CDM. However, when DESI BAO are combined with external CMB and type Ia supernova datasets, several analyses exhibit a mild preference for evolving dark energy in simple $w_0 w_a$ models, often with $w_0$ close to $-1$ and $w_a < 0$, corresponding to a transient phantom-like excursion in $w(z)$ over intermediate redshifts while remaining compatible with $\Lambda$CDM within current uncertainties (\cite{DESI2024VI}). It is important to emphasize that this is not yet a definitive detection of dynamical dark energy, and its quantitative significance depends on the adopted parametrization and data treatment. In this work we therefore use the DESI-motivated CPL behaviour as an illustrative benchmark trajectory for transient dynamics rather than as a claim of discovery. 
\\ \\
Our goal is to provide a minimal and calculable mechanism in which a transient departure from $w = -1$ arises as an effective macroscopic response of space itself. We model space as a three dimensional brane with a uniform background tension that plays the role of a bare cosmological constant and that admits a Nambu--Goto type description in the gravitational effective action (\cite{Goto1971,KhanTsDE}). On top of this background we introduce a phonon-like medium on the brane, described within the effective field theory of continuous media by a triplet of scalar fields that label comoving volume elements (\cite{Dubovsky2012, Endlich2013SolidInflation, NicolisRosen2015Elasticity}). 
\\ \\
In particular, Section 2 develops the elastic sector in a covariant and parameter-controlled way, identifying the small set of dimensionless combinations that govern the bulk modulus, longitudinal sound speed, and the background dark-energy-like response of the phonon medium. A central ingredient for obtaining a controlled transient phantom-like phase without introducing ghosts is dissipation and relaxation. Section 3 introduces a viscoelastic constitutive relation in which bulk viscous stress relaxes toward its instantaneous elastic value on a characteristic timescale, implemented via a Maxwell-type relaxation law motivated by the Kubo description of bulk viscous response (\cite{Kubo1957, Hosoya1984KuboCosmo, Jeon1995Transport, LandauFluid, Bemfica2022ViscousGR}). When the Hubble rate becomes comparable to the phonon gap and relaxation scales, the effective pressure can temporarily become more negative, producing an epoch in which $w_{\rm eff}(z)$ dips below $-1$ before returning toward a non-phantom value at both early and late times.
 \\ \\
Section 4 then connects this microphysical description to observations using a deliberately transparent compressed-likelihood strategy. Rather than re-running a full end-to-end likelihood analysis, we build a Gaussian approximation in the $(w_0,w_a)$ plane using the public DESI DR1 cosmology posterior products, specifically the Cobaya MCMC chains released for the DESI DR1 BAO cosmology analysis and its standard external dataset combinations (\cite{Planck2018, DESIDR1BAOChains, PantheonPlusData, PantheonPlusCosmo, Cobaya}). This yields a mean vector and covariance for $(w_0,w_a)$ that define a compact target for model comparison. In this sense our data interface is modular, easy to update when new DESI releases appear, and sufficient for establishing whether the viscoelastic phonon framework can naturally realize the type of transient CPL-like behaviour suggested by current combined analyses. 
\\ \\
The remainder of the paper develops the background cosmology implied by the elastic and viscous sectors, maps the theory parameters to an effective $w_{\rm eff}(z)$ trajectory, and quantifies the agreement with the DESI-motivated benchmark within the compressed likelihood. We also summarize the physical interpretation of the key parameter combinations and outline how the framework can be extended to more complete perturbation-level tests and to future data releases.

\section{\centering The Bulk Modulus of Space}

\subsection{\centering Scalar Description of Longitudinal Spatial Phonons}

In (\cite{KhanTsDE}), it was proposed that physical space can be modelled as a fundamental elastic three brane. The brane is described by a Nambu--Goto (\cite{Goto1971, Polchinski1998}) type worldvolume action,

\begin{equation}
	S_s \;=\; - T_s \int d^4 x \, \sqrt{-g}
\end{equation}
where $T_s$ is the geometric tension of space and $g_{\mu\nu}$ is the spacetime metric. In the same work (\cite{KhanTsDE}), the `no geometric sequester theorem' was formulated. The theorem states that in any minimal matter vacuum sequester (\cite{KaloperPadilla2014PRD, KaloperPadilla2014PRL}) the global constraint cancels constant contributions that arise from the matter sector but leaves a purely geometric unit operator $\int d^4 x \sqrt{-g}$ untouched. As a result the coefficient $T_s$ cannot be neutralized by the matter sequester and remains as a residual geometric contribution after renormalization from graviton loops and matter--graviton loops. This residual geometric tension can then be identified with the observed dark energy density at some scale $\mu_{\rm IR}$ appropriate for cosmology.
\\ \\
Nevertheless, $S_s$ controls only the uniform tension and does not describe how space responds to local compressions or rarefactions. To capture the elastic response we now introduce additional fields that live on the brane and describe longitudinal phonons or branons. The idea is that space is a continuous medium and we can attach internal labels to each infinitesimal element of this medium. These labels are represented by scalar fields.
\\ \\
In particular, we introduce three scalar fields (\cite{Dubovsky2012, Endlich2013SolidInflation, NicolisRosen2015Elasticity}),

\begin{equation}
	\phi^I(x) \, , \qquad I = 1, 2, 3
\end{equation}
which are defined on the four dimensional spacetime manifold. At each spacetime point $x^\mu$ the three numbers $\phi^I(x)$ provide an internal coordinate that labels which ``piece'' of the brane passes through that point. The fields are scalars under spacetime diffeomorphisms. They do not introduce new spatial dimensions. They are simply bookkeeping devices that track the configuration of the brane.
\\ \\
In the unperturbed configuration where space is homogeneous and isotropic we choose,

\begin{equation}
	\phi^I_{\text{bg}}(x) = x^I
\end{equation}
in suitable coordinates. This means that the internal labels coincide with ordinary comoving spatial coordinates. Small departures from this configuration can be expressed as,

\begin{equation}
	\phi^I(x) = x^I + \pi^I(x)
\end{equation}
which describe displacements of fluid elements of space. The fields $\pi^I$ describe longitudinal and transverse phonons. Our analysis focuses on the longitudinal sector as this is the sector that controls the bulk modulus and the bulk viscosity. Transverse phonons arise only when the brane is treated as embedded in a higher dimensional space, and we do not adopt that geometric picture here.
\\ \\
From the three scalars we build the standard invariant,

\begin{equation}
	B^{IJ} \equiv g^{\mu\nu} \, \partial_\mu \phi^I \, \partial_\nu \phi^J
\end{equation}
This object is a $3 \times 3$ matrix in the internal indices $I,J$. Under spacetime diffeomorphisms the scalars $\phi^I$ transform as ordinary scalar fields and the metric transforms as a rank two tensor. Therefore $B^{IJ}$ is a spacetime scalar and depends only on the internal indices. If we require that the medium is isotropic in the internal space and has no preferred directions then any local scalar built from $B^{IJ}$ must be invariant under internal $\mathrm{SO}(3)$ rotations,

\begin{equation}
	\phi^I \to R^I_{\ J} \, \phi^J \, , \qquad R \in \mathrm{SO}(3)
\end{equation}
A convenient isotropic invariant is then,

\begin{equation}
	b \equiv \sqrt{\det B^{IJ}}
\end{equation}
The quantity $b$ is a scalar under spacetime diffeomorphisms and also invariant under internal rotations of $\phi^I$. In the homogeneous background $\phi^I = x^I$ and in Minkowski spacetime one finds $B^{IJ} = \delta^{IJ}$ and therefore $b = 1$. If the configuration is compressed or rarefied then $B^{IJ}$ changes and $b$ departs from its background value. Intuitively $b$ measures the local comoving number density of brane elements or equivalently the inverse of the local volume per element. Large $b$ corresponds to compression and small $b$ corresponds to rarefaction.
\\ \\
With this invariant we can write the most general low energy action for longitudinal phonons that respects diffeomorphism invariance and internal isotropy as,

\begin{equation}
	S_{\text{ph}} = \int d^4x \, \sqrt{-g} \, F(b)
\end{equation}
where $F$ is an arbitrary function. The total action for space is then,

\begin{equation}
	S_{\text{space}} = S_s + S_{\text{ph}} = -T_s \int d^4 x \sqrt{-g}	+ \int d^4x \sqrt{-g} \, F(b) 
\end{equation}
The first term encodes the uniform geometric tension of space that survives sequestering. The second term encodes the elastic response of space through the scalar fields. Together they describe a medium that has both tension and phonon excitations.
\\ \\
The stress tensor of the phonon sector follows from varying $S_{\text{ph}}$ with respect to the metric which we obtain below.

\subsection{\centering Stress-Energy Tensor of the Phonon Fluid}

The stress energy tensor is defined in the standard way by (\cite{Carroll:2004}),

\begin{equation}
	T_{\mu\nu}^{\text{ph}}	\equiv - \frac{2}{\sqrt{-g}} \, \frac{\delta S_{\text{ph}}}{\delta g^{\mu\nu}} 
\end{equation}
We first vary the action, which gives,

\begin{equation}
	\delta S_{\text{ph}} =	\int d^4x \, \left[\delta(\sqrt{-g}) \, F(b)
	+ \sqrt{-g} \, F_b(b) \, \delta b \right] 
	\label{delta S phonon}
\end{equation}
where $F_b(b) \equiv dF/db$. The variation of the determinant of the metric is,

\begin{equation}
	\delta(\sqrt{-g}) = - \frac{1}{2} \sqrt{-g} \, g_{\mu\nu} \, \delta g^{\mu\nu}
	\label{delta g}
\end{equation}
The remaining task is therefore to compute $\delta b$ in terms of $\delta g^{\mu\nu}$. Note that the matrix $B^{IJ}$ depends on the metric through,

\begin{equation}
	B^{IJ} = g^{\mu\nu} \, \partial_\mu\phi^I \, \partial_\nu\phi^J
\end{equation}
so its variation is,

\begin{equation}
	\delta B^{IJ} =	\delta g^{\mu\nu} \, \partial_\mu\phi^I \, \partial_\nu\phi^J 
\end{equation}
Also,
\begin{equation}
	b^2 = \det B 
\end{equation}
The variation of the determinant is,

\begin{equation}
	\delta(\det B)	= (\det B) \, (B^{-1})_{JI} \, \delta B^{IJ} 
\end{equation}
where $(B^{-1})_{IJ}$ is the matrix inverse of $B^{IJ}$. Since $b = (\det B)^{1/2}$ we have,

\begin{equation}
	\delta b = \frac{1}{2} b^{-1} \, \delta(\det B) = \frac{1}{2} b^{-1} \, (\det B) \, (B^{-1})_{JI} \, \delta B^{IJ}	= \frac{1}{2} b \, (B^{-1})_{JI} \, \delta B^{IJ}
\end{equation}
where in the last step we used $\det B = b^2$. Substituting $\delta B^{IJ}$ gives,

\begin{equation}
	\delta b =  \frac{1}{2} b \, (B^{-1})_{JI} \left( \delta g^{\mu\nu} \, \partial_\mu\phi^I \, \partial_\nu\phi^J \right) = \frac{1}{2} b \, (B^{-1})_{IJ} \, \partial_\mu\phi^I \, \partial_\nu\phi^J \, \delta g^{\mu\nu} 
	\label{delta b}
\end{equation}
We now insert \eqref{delta g} and \eqref{delta b} into \eqref{delta S phonon} and collect the terms that multiply $\delta g^{\mu\nu}$, which gives,

\begin{align}
	\delta S_{\text{ph}} & = \int d^4x \, \left[ - \frac{1}{2} \sqrt{-g} \, g_{\mu\nu} \, F(b) \, \delta g^{\mu\nu}	+ \sqrt{-g} \, F_b(b) \,
	\frac{1}{2} b \, (B^{-1})_{IJ} \partial_\mu\phi^I \partial_\nu\phi^J \,
	\delta g^{\mu\nu} \right] \notag \\
	& =	\frac{1}{2} \int d^4x \, \sqrt{-g}	\left[ - g_{\mu\nu} F(b)
	+ b F_b(b) \, (B^{-1})_{IJ} \partial_\mu\phi^I \partial_\nu\phi^J
	\right] \delta g^{\mu\nu}
\end{align}
Comparing this with the definition of the stress energy tensor we obtain,

\begin{equation}
	T_{\mu\nu}^{\text{ph}} = F(b) \, g_{\mu\nu}	- b F_b(b) \, (B^{-1})_{IJ} \, \partial_\mu\phi^I \, \partial_\nu\phi^J 
	\label{stress energy tensor of phonon}
\end{equation}
The standard form of a perfect fluid stress tensor is,

\begin{equation}
	T_{\mu\nu}^{\text{pf}} = \big(\rho_{\text{ph}} + p_{\text{ph}}\big) \, u_\mu u_\nu	+ p_{\text{ph}}\, g_{\mu\nu}
	\label{normal se fluid tensor}
\end{equation}
To match $T_{\mu\nu}^{\text{ph}}$ to this form we express everything in terms of $g_{\mu\nu}$ and $u_\mu u_\nu$. We start by defining the identically conserved current,

\begin{equation}
	J^\mu \equiv \frac{1}{6} \, \epsilon^{\mu\nu\rho\sigma} \, \epsilon_{IJK} \, \partial_\nu\phi^I \, \partial_\rho\phi^J \, \partial_\sigma\phi^K
\end{equation}
where $\epsilon^{\mu\nu\rho\sigma}$ is the Levi–Civita symbol and $\epsilon_{IJK}$ is the antisymmetric symbol in the internal space. The conserved current satisfies $\nabla_\mu J^\mu = 0$ as a consequence of antisymmetry (see Appendix A). Its norm is (see Appendix B),

\begin{equation}
	J_\mu J^\mu = - b^2
\end{equation}
so we can define the fluid four velocity by,

\begin{equation}
	u^\mu \equiv \frac{J^\mu}{b} \, , \qquad u_\mu u^\mu = -1
\end{equation}
The projector onto spatial directions orthogonal to $u^\mu$ is,

\begin{equation}
	h_{\mu\nu} \equiv g_{\mu\nu} + u_\mu u_\nu 
\end{equation}
One can show that the combination built from the scalar fields equals this projector (see Appendix C),

\begin{equation}
	h_{\mu\nu} \equiv (B^{-1})_{IJ} \, \partial_\mu\phi^I \partial_\nu\phi^J
	= g_{\mu\nu} + u_\mu u_\nu
	\label{projector formula main text}
\end{equation}
With this identification the stress tensor can be written as,

\begin{equation}
	T_{\mu\nu}^{\text{ph}}	= F(b)\, g_{\mu\nu} - b F_b(b) \, h_{\mu\nu}
\end{equation}
Using \eqref{projector formula main text} in \eqref{stress energy tensor of phonon} we get,

\begin{align}
	T_{\mu\nu}^{\text{ph}} & = F(b) \, g_{\mu\nu} - bF_b(b) \big(g_{\mu\nu} + u_\mu u_\nu\big) \notag \\[4pt]
	& = \big(F(b) - bF_b(b)\big) \, g_{\mu\nu} - bF_b(b) \, u_\mu u_\nu 
\end{align}
Comparing with \eqref{normal se fluid tensor}, we can immediately read off,

\begin{equation}
	p_{\text{ph}}(b) = F(b) - bF_b(b) \, , \qquad \rho_{\text{ph}}(b) + p_{\text{ph}}(b) = - bF_b(b) 
	\label{pressure of phonon fluid}
\end{equation}
Solving the second relation using the first gives,

\begin{equation}
	\rho_{\text{ph}}(b)	= - F(b) 
	\label{energy density of phonon fluid}
\end{equation}
Equivalently,

\begin{equation}
	\rho_{\text{ph}}(b) = u^\mu u^\nu T_{\mu\nu}^{\text{ph}} \, , \qquad
	p_{\text{ph}}(b) = \frac{1}{3}\,h^{\mu\nu} T_{\mu\nu}^{\text{ph}}
\end{equation}
which reproduce the same results when $T_{\mu\nu}^{\text{ph}}$ is inserted. In Sec.~2.3, we will identify the bulk modulus of space by using phonon physics at a background value $b_0$. We will verify the result obtained for bulk modulus in Sec.~2.5 by doing a background Taylor expansion at $b_0$.

\subsection{\centering Calculating the Spatial Bulk Modulus} 
The adiabatic sound speed of longitudinal phonons is defined by (see (\cite{WeinbergCosmology}) for fluids in FLRW and sound speed discussions),

\begin{equation}
	c_s^2 \equiv \frac{d p_{\text{ph}}}{d \rho_{\text{ph}}}
	= \frac{\dfrac{d p_{\text{ph}}}{d b}}{\dfrac{d \rho_{\text{ph}}}{d b}}
\end{equation}
where the derivatives are taken at fixed entropy per comoving element. From the expressions \eqref{pressure of phonon fluid} and \eqref{energy density of phonon fluid} above we find,

\begin{align}
	\frac{d \rho_{\text{ph}}}{d b} & = - F_b(b), \quad \text{where } F_b = \frac{dF}{db} \notag \\[4pt]
	\frac{d p_{\text{ph}}}{d b}	& = F_b(b) - \Bigl[F_b(b) + b F_{bb}(b)\Bigr]	= - b F_{bb}(b), \quad \text{where } F_{bb} = \frac{d^2F}{db^2}
\end{align}
Therefore,

\begin{equation}
	c_s^2 = \frac{d p_{\text{ph}}/d b}{d \rho_{\text{ph}}/d b} = \frac{- b F_{bb}(b)}{- F_b(b)} = \frac{b F_{bb}(b)}{F_b(b)} 
	\label{phonon speed equation}
\end{equation}
In a relativistic fluid, the adiabatic bulk modulus is given as,

\begin{equation}
	K_{\text{ph}} \equiv \left(\rho_{\text{ph}} + p_{\text{ph}}\right) c_s^2
\end{equation}
which can be verified directly from the relation between pressure and energy density under a homogeneous compression at fixed entropy. Combining this with the expressions above gives,

\begin{equation}
	K_{\text{ph}} = \left[\rho_{\text{ph}}(b) + p_{\text{ph}}(b)\right] c_s^2 = \left[- b F_b(b)\right] \frac{b F_{bb}(b)}{F_b(b)} = - b^2 F_{bb}(b)
	\label{relation between K and c_s}
\end{equation}
By convention the bulk modulus is positive, so for stability one requires $F_{bb}(b_0) < 0$ at the background value $b_0$. Therefore, we have,

\begin{equation}
	K_{\text{ph}} = - b_0^2 F_{bb}(b_0)
\end{equation}
Equation \eqref{phonon speed equation} however dictates that $F_{bb}(b_0) < 0$ translates to $c_s^2 < 0$, which is non physical. However, at the same time note that we also require,

\begin{equation}
	\rho_{\text{ph}}(b_0) + p_{\text{ph}}(b_0) > 0
\end{equation}
Using \eqref{pressure of phonon fluid} in the equation above gives,

\begin{equation}
	- b_0 \, F_b (b_0) > 0 \quad \Longrightarrow \quad b_0 \, F_b (b_0) < 0
\end{equation}
 and since $b_0 >0$, it must then be true that,
 
 \begin{equation}
 	 F_b (b_0) < 0
 \end{equation}
 which implies $c_s^2$ is in fact positive for the background $b_0$ because,
 
 \begin{equation}
 	c_s^2 = \frac{b_0 F_{bb}(b_0)}{F_b(b_0)}, \quad b_0 > 0, \quad F_{bb}(b_0) < 0, \quad F_b (b_0) <0
 \end{equation}
With this in mind, we now parameterize the stiffness of space in terms of the geometric tension,

\begin{equation}
	K_{\text{ph}} \equiv \kappa \, T_s
\end{equation}
where $\kappa$ is a dimensionless parameter that measures how stiff the phonon medium is compared to the background tension. Note using equation \eqref{relation between K and c_s}, we can express the longitudinal speed of phonons as,

\begin{equation}
	c_s^2 = \frac{K_{\rm ph}}{{\rho_{\text{ph}}(b) + p_{\text{ph}}(b)}}
\end{equation} 
Now let us assume the following as a working ansatz,

\begin{equation}
	\rho_{\text{ph}}(b) + p_{\text{ph}}(b) \equiv \rho_{\rm eff} = \varepsilon \, T_s, \quad \text{where } 0 < |\varepsilon| < 1
 \end{equation}
which gives the phonon speed as,
 
 \begin{equation}
 	c_s^2 = \frac{K_{\rm ph}}{{\rho_{\text{ph}}(b) + p_{\text{ph}}(b)}} = \frac{\kappa \, T_s}{\varepsilon \, T_s} = \frac{\kappa}{\varepsilon}
 \end{equation}
Causality and stability require $c_s^2 \le 1$. This translates into a bound,

\begin{equation}
	0 < \frac{\kappa}{\varepsilon} \le 1
	\label{phonon speed constrain}
\end{equation} 
where $\kappa/\varepsilon = 1$ corresponds to luminal longitudinal phonons and smaller values correspond to softer subluminal propagation. We therefore have two free parameters $\kappa$ and $\varepsilon$ such that their ratio should satisfy the constrain \eqref{phonon speed constrain}.

\subsection{\centering A Caveat for the Phonon Vacuum Energy Density}
We start from the most general form,

\begin{equation}
	S_{\text{space}} = - T_s\int d^4x \sqrt{-g}	+ \int d^4x \sqrt{-g} \, F(b)
\end{equation}
where both $T_s$ and $F(b)$ are understood as renormalized quantities. We expand around a homogeneous equilibrium configuration $b = b_0 + \delta b$ and write,

\begin{equation}
	F(b) = F(b_0) + F_b(b_0) \, \delta b + \frac{1}{2} F_{bb}(b_0) \, \delta b^2 + \ldots
	\label{expansion of F}
\end{equation}
The constant term $F(b_0)$ multiplies $\sqrt{-g}$ and therefore behaves as a contribution to the cosmological constant. It is convenient to absorb this constant into the geometric tension. We define,

\begin{equation}
	T_s^{\text{eff}} \equiv T_s + F(b_0), \qquad \tilde F(b) \equiv F(b) - F(b_0)
	\label{absorbtion convention}
\end{equation}
In terms of these variables the action becomes,

\begin{equation}
	S_{\text{space}} = - T_s^{\text{eff}} \int d^4x \sqrt{-g} + \int d^4x \sqrt{-g} \,\tilde F(b)
\end{equation}
and by construction $\tilde F(b_0) = 0$. Note that the full stress tensor is unchanged by this redefinition. We have only reshuffled a constant between the Nambu--Goto part and the phonon part, so all physical predictions that depend on the total $T_{\mu\nu}$ are the same. The no geometric sequester theorem in minimal sequesters then applies to $T_s^{\text{eff}}$ and states that this geometric unit operator is not cancelled by the matter constraint and it survives as residual dark energy after receiving loop corrections from graviton and matter-graviton loops. 
\\ \\
From this point on we drop the tilde and rename,

\begin{equation}
	T_s \equiv T_s^{\text{eff}}, \qquad	F(b) \equiv \tilde F(b)
	\label{renaming convention}
\end{equation}
with the renormalization condition,

\begin{equation}
	F(b_0) = 0
\end{equation}
With this convention the phonon sector has no vacuum energy at the equilibrium point,

\begin{equation}
	\rho_{\text{ph}}(b_0) = - F(b_0) = 0
\end{equation}
while the dominant dark energy density is carried entirely by the geometric tension $T_s$. The elastic response and the deviation from a pure cosmological constant depend on derivatives of $F$ evaluated at $b_0$, for example,

\begin{equation}
	p_{\text{ph}}(b_0) = - b_0 F_b(b_0), \qquad	K_{\text{ph}} = - b_0^2 F_{bb}(b_0), \qquad	c_s^2 = \frac{b_0 F_{bb}(b_0)}{F_b(b_0)}
\end{equation}
which are unaffected by the constant shift.

\subsection{\centering Effective Potential $F(b)$ Near Equilibrium}

The previous discussion expressed the phonon fluid in terms of an arbitrary
function $F(b)$ and related its derivatives at the equilibrium point $b_0$ to the physical quantities $\rho_{\rm ph}$, $p_{\rm ph}$ and $K_{\rm ph}$. Furthermore, for phenomenological purposes we introduced the parameters $\varepsilon$ and $\kappa$ through,

\begin{equation}
	\rho_{\rm ph}(b_0) = 0, \qquad \rho_{\rm ph}(b_0) + p_{\rm ph}(b_0) \equiv \varepsilon T_s,	\qquad K_{\rm ph} \equiv \kappa T_s
\end{equation}
with $0 < \varepsilon < 1$ and $0 < \kappa < 1$. Using the relations,

\begin{equation}
	\rho_{\rm ph}(b) = -F(b), \qquad \rho_{\rm ph}(b) + p_{\rm ph}(b) = -b F_b(b),	\qquad K_{\rm ph} = -b^2 F_{bb}(b)
\end{equation}
these conditions become,

\begin{equation}
	F(b_0) = 0, \qquad -b_0 F_b(b_0) = \varepsilon T_s, \qquad -b_0^2 F_{bb}(b_0) = \kappa T_s
\end{equation}
A simple class of functions that satisfies all three relations and the stability requirements $F_b(b_0) < 0$ and $F_{bb}(b_0) < 0$ is obtained by expanding $F(b)$ to quadratic order,
\begin{equation}
	F(b) = - F_b (b_0) \left(b - b_0 \right) - \frac{F_{bb}(b_0)}{2} \left(b - b_0 \right)^2 + \mathcal{O} \left(b-b_0 \right)^3
	\label{eq:Fb-example}
\end{equation}
where as before $F_b = F'(b) = \dfrac{dF(b)}{db}$ and similarly $F_{bb} = F''(b) = \dfrac{d^2F(b)}{db^2}$. Taking $b_0$ common, we find,

\begin{equation}
	F(b) = - F_b(b_0) \, b_0 \, \left(\frac{b}{b_0} - 1\right) - \frac{F_{bb}(b_0)}{2} \, b_0^2 \left(\frac{b}{b_0} - 1 \right)^2 + \mathcal{O} \left(b - b_0 \right)^3
	\label{eq:Fb-example}
\end{equation}
Let,

\begin{equation}
	 F_b(b_0) \, b_0 = - \varepsilon \, T_s \quad \text{and} \quad \frac{F_{bb}(b_0)}{2} \, b_0^2 = - \kappa \, T_s
\end{equation}
At $b = b_0$ one finds,

\begin{equation}
	F(b_0) = 0, \qquad F_b(b_0) = -\frac{\varepsilon T_s}{b_0} < 0,	\qquad
	F_{bb}(b_0) = -\frac{\kappa T_s}{b_0^2} < 0
\end{equation}
so that
\begin{equation}
	\rho_{\rm ph}(b_0) = 0, \qquad \rho_{\rm ph}(b_0) + p_{\rm ph}(b_0) = \varepsilon T_s, \qquad	K_{\rm ph} = \kappa T_s
\end{equation}
and the adiabatic sound speed is,

\begin{equation}
	c_s^2 = \frac{K_{\rm ph}}{\rho_{\rm ph} + p_{\rm ph}} = \frac{\kappa}{\varepsilon}
\end{equation}
This example shows explicitly that the phenomenological parameters $\varepsilon$ and $\kappa$ which we will use in Secs.~3 and 4 can be viewed as the first two derivatives of a local potential $F(b)$ evaluated at the equilibrium configuration $b_0$. Higher order terms in \eqref{eq:Fb-example} are not needed for the small homogeneous deformations that are relevant for the late time cosmological evolution considered in this work. We keep $\varepsilon$ and $\kappa$ as phenomenological parameters as mentioned before. We not derive them here from first principles, reserving that analysis for a future work.

\section{\centering The Bulk Viscosity of Space}
In this section we discuss how the elastic three brane that models space can acquire a non zero bulk viscosity and how this viscous response can decay as the universe expands. The starting point is the phonon fluid of Sec.~2, described by the scalar invariant $b$ and the function $F(b)$. The Nambu-Goto term contributes a constant geometric tension $T_s$ and the phonon sector captures small compressions and rarefactions around the equilibrium value $b_0$.

\subsection{\centering Microscopic definition of bulk viscosity}

In a relativistic fluid the bulk viscosity describes the dissipative response of the pressure to a homogeneous expansion or contraction. In covariant form the viscous part of the stress tensor can be written as (see (\cite{LandauFluid,Bemfica2022ViscousGR,  Weinberg1971BulkViscosity, Zimdahl2001BulkViscous, Colistete2007BulkViscous, Kovtun2012Lectures}) for related discussion),

\begin{equation}
	T^{\mu\nu}_{\text{visc}} = - \zeta \, (\nabla_\alpha u^\alpha) \, h^{\mu\nu}
	\label{viscosity tensor}
\end{equation}
where $u^\mu$ is the fluid four velocity and $h^{\mu\nu} = g^{\mu\nu} + u^\mu u^\nu$ as defined earlier. 
\\ \\
The microscopic definition of $\zeta$ can be given through a Kubo formula in terms of the retarded correlator of the stress tensor (\cite{Kubo1957, Hosoya1984KuboCosmo, Jeon1995Transport}). A convenient expression is,

\begin{equation}
	\zeta = \frac{1}{9} \lim_{\omega \to 0} \frac{1}{\omega} \operatorname{Im} G^R_{T^i_{\ i} T^j_{\ j}}(\omega, \mathbf{k} = 0)
\end{equation}
where,

\begin{equation}
	G^R_{T^i_{\ i} T^j_{\ j}}(\omega, \mathbf{k}) = - i \int d^4x \, e^{i\omega t - i \mathbf{k} \cdot \mathbf{x}} \theta(t)	\left\langle \left[ T^i_{\ i}(t, \mathbf{x}), T^j_{\ j}(0, \mathbf{0}) \right] \right\rangle_{T}
\end{equation}
is the retarded Green function in a thermal state at temperature $T$. The trace $T^i_{\ i}$ is taken over spatial indices. The imaginary part of this correlator measures dissipation into microscopic degrees of freedom during a slow compression.
\\ \\
In the elastic brane picture the total stress tensor is
\begin{equation}
	T_{\mu\nu} = - T_s g_{\mu\nu} + T^{\text{ph}}_{\mu\nu}
\end{equation}
The Nambu-Goto term has no dynamics and contributes only a constant. It does not contribute to the imaginary part of the stress correlator. The frequency dependent and dissipative part of \(G^R_{T^i_{\ i} T^j_{\ j}}\) is entirely due to the phonon sector. In the Kubo formula for \(\zeta\) we can therefore replace $T^i_{\ i}$ by the trace of the phonon stress tensor.

\subsection{\centering Physical picture on the Elastic Brane}

We now discuss qualitatively how the phonon sector can produce a bulk viscosity and why this viscosity falls as the universe expands. The elastic brane supports longitudinal phonons. These are collective excitations of the scalar fields $\phi^I$ that describe compressions and rarefactions of space. At the linearized level one finds a gapless mode with dispersion $\omega^2 = c_s^2 k^2$. In a more complete theory higher derivative operators, loop corrections and possible small sources of explicit symmetry breaking can modify this relation and can generate a small mass gap,
\begin{equation}
	\omega^2 \simeq c_s^2 k^2 + m_\phi^2
\end{equation}
for the longitudinal excitation. In this work we assume that such microscopic physics fixes a non zero but small gap $m_\phi$ for the relevant phonon mode. We do not derive $m_\phi$ from first principles here. Instead we treat $m_\phi$ as a phenomenological parameter that can be constrained by cosmological observations. The gap is assumed to be much smaller than the microscopic cutoff of the effective theory and much smaller than the intrinsic scale set by the tension of space, so that the fluid description remains valid.
\\ \\
In this work, we will interpret the Gibbons-Hawking temperature (\cite{GibbonsHawking1977}) of a spatially flat FRW spacetime,

\begin{equation}
	T_H \equiv \frac{H}{2\pi}
	\label{eq:GHtemp}
\end{equation}
as an \emph{effective} horizon temperature associated with the cosmological
apparent horizon, rather than as literal Hawking radiation from an exact
de~Sitter event horizon. This viewpoint follows the horizon-thermodynamics
approach in which the Friedmann equations can be written as a first law
$dE = T_H dS + W dV$ at the apparent horizon, with $T_H = 1/2\pi R_A$
and $R_A = 1/H$ for $k = 0$ FRW backgrounds (\cite{AkbarCai2007, CaiCao2007, GongWang2007, Sheykhi2010}). We assume that the phonon sector is in near-equilibrium with this effective thermal bath at temperature $T_H$ during the epoch when bulk viscosity is relevant. Concretely, we assume that the microscopic processes that create and annihilate phonons and that redistribute their momentum are fast enough that the phonon distribution relaxes toward a local equilibrium characterized by $T_H$ on time scales shorter than or comparable to the Hubble time during the epoch when bulk viscosity is relevant. Under this assumption the Kubo formula and the Maxwell viscoelastic model provide a meaningful characterization of the bulk viscosity $\zeta$ in terms of correlation functions evaluated at temperature $T_H$. We will discuss these ideas in the Sec.~3.3.
\\ \\
If the phonon mass gap is of some related infrared scale, the horizon temperature decreases relative to this gap as the universe expands. Once the horizon temperature falls below the gap, $T_H \ll m_\phi$, real phonon excitations are strongly suppressed. Scattering rates become very small and stress tensor fluctuations relax only on very long timescales. In this regime the imaginary part of the stress correlator, and hence $\zeta(T_H)$, are expected to be suppressed by a Boltzmann factor of the form,

\begin{equation}
	\zeta(T_H) \propto e^{- m_\phi / T_H}
\end{equation}
up to powers of temperature. This behaviour means that the viscous response is important only during an intermediate epoch when the horizon temperature is high enough to excite phonons. As the universe cools and $T_H$ drops, the bulk viscosity decays rapidly and space behaves as an almost perfect elastic medium with negligible dissipation as we discuss at length in the discussion below.
\\ \\
Most crucially, we interpret the phonon sector as describing small elastic deformations of the worldvolume of space. The bulk modulus $K_{\rm ph}$ and the bulk viscosity $\zeta$ are therefore understood as elastic and dissipative response coefficients of the brane of space itself, rather than of an additional matter fluid. We do not introduce any new long lived matter components beyond the standard cosmological fluids. The only new degrees of freedom are the phonons that propagate on the elastic worldvolume and encode its response to compression and expansion. 

\subsection{\centering Maxwell Visco-Elastic Model and Cosmological Bulk Viscosity in FLRW}

The Kubo formula defines $\zeta$ in the limit of very low frequency compared to microscopic scales. For cosmological evolution the relevant frequency is of order the Hubble parameter $H$. It is therefore useful to introduce an effective bulk viscosity $\zeta_{\text{eff}}(H)$ that encodes both the microscopic relaxation time and the finite timescale of the expansion.
\\ \\
For a spatially flat FLRW metric (\cite{Mukhanov:2005}), with scale factor $a(t)$ and comoving fluid, one has $\nabla_\alpha u^\alpha = 3 H$, where $H = \dot a / a$ is the Hubble parameter. The bulk viscous pressure $\Pi$ is then the spatial trace of equation \eqref{viscosity tensor}, which gives,

\begin{equation}
	\Pi = - 3 \zeta H
	\label{phonon pressure}
\end{equation}
Note that a positive bulk viscosity produces a negative $\Pi$ during expansion (\cite{Zimdahl2001BulkViscous, Colistete2007BulkViscous, Singh2008BulkViscous, Gariel1995BulkViscous, Czajka2019BulkQCD}).
\\ \\
A simple and physically transparent way to obtain $\zeta$ is provided by a Maxwell visco-elastic model for the bulk stress (\cite{Maxwell1867}). We assume that the bulk viscous pressure $\Pi$ obeys the simple linear causal relationship,

\begin{equation}
	\tau(H) \, \dot{\Pi} + \Pi = - 3 \, \zeta_0(H) \, H
	\label{Maxwell Equation}
\end{equation}
where $\tau(H)$ is a relaxation time.
\\ \\
Consider now a homogeneous perturbation with frequency $\omega$ of the medium. Fourier transforming the Maxwell equation \eqref{Maxwell Equation} gives the left hand side as,

\begin{equation}
	\tau(H) \, \dot{\Pi}(t) + \Pi(t) = \int \frac{d\omega}{2\pi} [(1 -i \omega \tau) \Pi (\omega) ]e^{-i \omega t}
\end{equation}
and the right hand side as,

\begin{equation}
	-3 \zeta_0 H(t) = \int \frac{d\omega}{2\pi} [-3 \zeta_0 H(\omega)] e^{-i\omega t}
\end{equation}
Since the two equations above are equal for all $t$, the integrands must agree frequency by frequency. This gives the algebraic relation,

\begin{equation}
	(1 - i \omega \tau) \, \Pi(\omega) = - 3 \, \zeta_0 \, H(\omega)
	\label{Complex Maxwell Equation}
\end{equation}
It is natural to define an effective bulk viscosity at frequency \(\omega\) by,

\begin{equation}
	\Pi(\omega) = - 3 \zeta_{\text{eff}}(\omega) H(\omega)
\end{equation}
In a general viscoelastic medium the real and imaginary parts of $\zeta$ encode different aspects of the response. The imaginary part is related to reactive effects, while the real part controls energy dissipation. In a spatially flat Friedmann background the expansion is monotonic and there is no oscillatory driving with a sharply defined frequency. We are interested in the effective bulk viscosity that enters the average dissipated power and the background effective pressure. For these purposes it is natural to work with the real part of the complex Maxwell viscosity as obtained in equation \eqref{Complex Maxwell Equation}. This gives,

\begin{equation}
	\zeta_{\text{eff}}(\omega) = \frac{\zeta_0(H)}{1 + \omega^2 \tau(H)^2}
	\label{Maxwell Bulk Viscosity Real}
\end{equation}
We identify the combination $K_{\text{ph}} \tau(H)$ has the dimensions of viscosity and plays the role of the static bulk viscosity of the medium  (\cite{Roylance2001}). In a Maxwell model one typically has $\zeta_0(H) \sim K_{\text{ph}} \tau(H)$. This relation reflects the fact that the same microphysics that sets the elastic modulus also sets the relaxation time. Substituting this scaling relation gives,
 
\begin{equation}
	\zeta_{\text{eff}}(\omega) \simeq \frac{K_{\text{ph}} \tau(H)}{1 + \omega^2 \tau(H)^2}
\end{equation}
For cosmology the relevant frequency is the Hubble rate $\omega \sim H$. The effective bulk viscosity is therefore,

\begin{equation}
	\zeta_{\text{eff}}(H) \simeq \frac{K_{\text{ph}} \, \tau(H)}{1 + H^2 \tau(H)^2}
	\label{effective viscosity}
\end{equation}
This expression smoothly interpolates between two regimes of fast and slow relaxations as shown below. 
\\ \\
The relaxation time $\tau(H)$ is controlled by the inverse of the microscopic interaction rate $\Gamma (T)$ for phonons, that is,

\begin{equation}
	\tau \sim \frac{1}{\Gamma (T)}
\end{equation}
At temperature $T_H$, the interaction rate for processes that involve real phonon excitations typically behaves as (\cite{KolbTurner:1990}),

\begin{equation}
	\Gamma(T_H) \sim n_{\rm ph} (T_H) \, \sigma_{\rm ph} \, v_{\rm ph}(T_H)
\end{equation}
where $n_{\rm ph}$ is the phonon number density, $\sigma_{\rm ph}$ is an effective cross section and $v_{\rm ph}$ is a typical relative velocity. For a gapped excitation with mass $m_\phi$ at temperature $T_H$, the number density follows,

\begin{equation}
	n_{\rm ph} (T_H) \simeq g \left( \frac{m_\phi T_H}{2\pi} \right)^{3/2} e^{- m_\phi / T_H}
\end{equation}
where $g$ counts polarizations and internal degrees of freedom. The thermal group velocity of phonons $v_{\rm ph}$ in the regime $T_H < m_{\phi}$ (we will consider only this regime in the later Sections of the paper as well) is of order,

\begin{equation}
	v_{\rm ph} \sim c_s \, \sqrt{\frac{T_H}{m_\phi}}
\end{equation} 
Therefore, we have,

\begin{equation}
	\Gamma(T_H) \sim  g \, c_s \, \sigma_{\rm ph} \, \left( \frac{m_\phi T_H}{2\pi} \right)^{3/2} e^{- m_\phi / T_H}  \, \sqrt{\frac{T_H}{m_\phi}}
\end{equation}
which gives,

\begin{equation}
	\Gamma(T_H) \sim  \frac{g \, c_s}{(2 \, \pi)^{3/2}} \  \sigma_{\rm ph} \ T_H^2 \ m_\phi \ e^{- m_\phi / T_H}
	\label{Gamma equation} 
\end{equation}
The important feature however is the Boltzmann factor in the collision rate,

\begin{equation}
	\Gamma(T_H) \propto e^{- m_\phi / T_H}
\end{equation}
up to powers of $T_H$ and $\mathcal{O}(1)$ constants. As a result the relaxation time grows rapidly at low temperature,

\begin{equation}
	\tau(T_H) \sim \frac{1}{\Gamma(T_H)} \propto e^{+ m_\phi / T_H}
\end{equation}
and results in,

\begin{equation}
	H \, \tau(H) \gg 1
\end{equation}
so that the effective viscosity in equation in \eqref{effective viscosity} is strongly suppressed.
\\ \\
On the other hand at very high temperatures, $T_H \gg m_\phi$, the Boltzmann suppression is absent and the relaxation time is short. In this regime it is natural to presume that,

\begin{equation}
	H \, \tau(H) \ll 1 
\end{equation}
so the medium follows the expansion almost instantaneously and the denominator in \eqref{effective viscosity} is close to unity. 
\\ \\
It is important to mention here that the detailed behavior of $H \tau(H)$ as a function of redshift is model dependent, since it depends on the structure of $\sigma(T)$ and on the microscopic couplings of the phonon sector. For phenomenological purposes it is convenient to assume that there exists an intermediate epoch in which $H \tau(H)$ passes through unity,

\begin{equation}
	H \, \tau(H) \sim 1 \quad \text{for} \quad H \simeq H_\star 
\end{equation}
with $H_\star$ a characteristic Hubble scale before the onset of dark energy domination.
\\ \\
In this work we assume that the onset of dark energy domination (at redshift $z_{\rm DE}$) occurs when the Gibbons--Hawking temperature is already past (well below) the phonon gap,

\begin{equation}
	T_H(z_{\rm DE}) < m_\phi
\end{equation}
so that $H \tau(H)$ is larger than order unity at the beginning of dark energy domination. As the universe expands and $H$ decreases, the temperature $T_H$ drops and at much later times, we have $T_H \ll m_\phi$ so that the relaxation time is large $H \tau(H) \gg 1$ and the effective viscosity decays rapidly.
\\ \\
We may now obtain the expression for $\sigma$ using the fact that $H\tau \sim 1$ when $H \sim H_{\star}$. Equation \eqref{Gamma equation} gives,

\begin{equation}
	\tau(T_{H_{\star}}) \sim \frac{1}{\Gamma(T_{H_{\star}})} \sim \frac{(2 \, \pi)^{3/2}}{g \, c_s \, \sigma_{\rm ph}} \ \frac{1}{T_{H_{\star}}^2 \ m_\phi}  \ e^{m_\phi / T_{H_{\star}}}
\end{equation}
Using the Gibbons-Hawking relation $T_{H_{\star}} = H_{\star}/2\pi$ in the equation above we get,

\begin{equation}
	\tau(T_{H_{\star}}) \sim \frac{1}{\Gamma(T_{H_{\star}})} \sim \frac{(2 \, \pi)^{3/2}}{g \, c_s \, \sigma_{\rm ph}} \ \frac{(2 \, \pi)^2}{H_{\star}^2 \ m_\phi}  \ e^{2 \pi m_\phi /H_{\star}}
	\label{Tau equation}
\end{equation}
Multiplying both sides by $H_{\star}$ gives,

\begin{equation}
	H_{\star} \, \tau(T_{H_{\star}}) \sim \frac{(2 \, \pi)^{7/2}}{g \, c_s \, \sigma_{\rm ph} \, H_{\star} \ m_\phi}  \ e^{2 \pi m_\phi /H_{\star}}
\end{equation}
Since $H_{\star} \, \tau \sim 1$, we have,

\begin{equation}
	\sigma_{\rm ph}\sim  \frac{(2 \, \pi)^{7/2}}{g \, c_s \, {H_{\star}\ m_\phi}}  \ e^{2 \pi m_\phi /H_{\star}}
\end{equation}
The mass gap is naturally defined as $m_\phi \equiv H_{\star}/2\pi$ or equivalently $H_{\star} \equiv 2 \, \pi m_\phi$, which gives,

\begin{equation}
	\sigma_{\rm ph} \sim  \frac{(2 \, \pi)^{9/2}}{g \, c_s \, H_{\star}^2}  \ e
	\label{expression for sigma}
\end{equation}
At first sight the scale implied by equation \eqref{expression for sigma} may appear very large since $\sigma_{\rm ph} \propto H_{\star}^{-2}$, the effective cross section is of order the horizon area when the viscous response is strongest. This does not signal any inconsistency because the phonons in our construction may not be considered as elementary particles propagating in vacuum with a microscopic two body scattering cross section. They are collective normal modes of the brane medium, which describe coherent distortions of many underlying degrees of freedom over horizon sized regions.
\\ \\
One encounters similar scenarios in ordinary condensed matter systems where phonons are also collective excitations. Their effective interaction length is set by the wavelength of the mode and by the correlation length of the medium rather than by an atomic size. The situation here is similar. The relevant scale in the bulk viscosity sector is the phonon Compton wavelength $\lambda_\phi \sim H_{\star}^{-1}$, which is comparable to the cosmological horizon when the viscous effects peak. An effective cross section of order $H_{\star}^{-2}$ simply reflects the fact that a given phonon mode samples and mixes the state of the medium over a patch whose linear size is set by $\lambda_\phi$.
\\ \\
The large value of \(\sigma_{\rm ph}\) therefore indicates that the phonon excitations are very efficient at redistributing stress within each horizon volume. It does not imply strong local scattering in the usual particle physics sense and does not lead to violations of bounds on microscopic self interactions. In the hydrodynamic limit the important quantity is the relaxation time that controls how fast the isotropic stress returns to equilibrium and this is already encoded in the Maxwell type viscosity model used in the previous sections. The interpretation of the phonons as long wavelength collective modes of the brane is consistent with an effective cross section of order the horizon area and with their role as infrared degrees of freedom that mediate the viscous response of dark energy.
\\ \\
With this in mind, we now substitute the result \eqref{expression for sigma} for $\sigma_{\rm ph}$ in equation \eqref{Tau equation} with the latter evaluated for a general $H$ and simultaneously use the definition $m_\phi \equiv H_{\star}/2\pi$ to arrive at,

\begin{equation}
	\tau (T_H) \sim \frac{(2 \, \pi)^{7/2}}{g \, c_s \, \sigma_{\rm ph}(T_H) \, H^2 \ m_\phi}  \ e^{2 \pi m_\phi /H} \sim  \frac{H_{\star}}{H^2}  \ e^{\frac{H_{\star}}{H} - 1}
\end{equation}
and we therefore get the dimensionless combination $H \tau$ as,

\begin{equation}
	H \tau(T_H) \sim \frac{H_{\star}}{H}  \ e^{\frac{H_{\star}}{H} - 1}
\end{equation}
For the purposes of this paper it is sufficient to treat $m_\phi$ and therefore $H_{\star}$ as phenomenological parameters and to require that the condition $T_{H_\star} \sim m_\phi$ is realized at some redshift $z_{\star}$ before the onset of dark energy domination. In essence, we assume that $H_\star$ is reached at some $z_\star > 0.3$, in fact, we may assume that $H_{\star}$ is reached even before cosmic acceleration starts, that is $z_\star > 0.6$. This ensures that the model produces a transient viscous epoch, with a mild phantom like deviation of the effective equation of state (see \cite{Caldwell2002Phantom} for a phantom like equation of state for dark energy) followed by a rapid return to an effectively non viscous or at least not significantly viscous dark energy. In Sec.~3.4 next, we will discuss the effective equation of state with the phonon pressure contribution and highlighting the phantom dip scenario.

\subsection{\centering Effective Equation of State of Space and the Phantom Dip}
The total dark energy sector consists of the geometric tension $T_s$ and the phonon fluid of Sec. 2. Using equations \eqref{absorbtion convention} and \eqref{renaming convention}, the homogeneous background energy density is simply the residual geometric tension,

\begin{equation}
	\rho_{\rm DE} \simeq T_s 
	\label{rhoDE-Ts}
\end{equation}
while the phonon sector contributes only a small enthalpy density and a dissipative component. 
\\ \\
In the absence of dissipation the pressure of the dark energy sector is then approximately,

\begin{equation}
	p_{\rm DE} \simeq - T_s + \rho_{\rm ph} + p_{\rm ph}
	\simeq - T_s + 0 + \varepsilon \, T_s \simeq - T_s + \varepsilon \, T_s
	\label{pDE-nonviscous}
\end{equation}
so the non viscous equation of state parameter is simply,

\begin{equation}
	w_{\rm DE} \equiv \frac{p_{\rm DE}}{\rho_{\rm DE}} \simeq -1 + \varepsilon
	\label{wDE-epsilon}
\end{equation}
which is very close to $-1$ for small $|\varepsilon|$. This reflects the fact that the dominant contribution to dark energy is the geometric tension $T_s$ that survives sequestering, while the phonon sector provides only a small correction to the pressure through its elastic response.
\\ \\
Once bulk viscosity is included the pressure that enters the Friedmann equations is modified by the viscous contribution $\Pi$, where,
\begin{equation}
	\Pi = - 3 \, \zeta_{\rm eff}(H) \, H 
	\label{Pi-def}
\end{equation}
where $\zeta_{\rm eff}(H)$ is the effective real bulk viscosity obtained from the Maxwell viscoelastic model in equation \eqref{effective viscosity}. The effective pressure of space is then,

\begin{equation}
	p_{\rm eff} = p_{\rm DE} + \Pi	= p_{\rm DE} - 3 \, \zeta_{\rm eff}(H) \, H 
	\label{peff-def}
\end{equation}
Using \eqref{rhoDE-Ts} and \eqref{pDE-nonviscous} the effective equation of state parameter becomes,

\begin{equation}
	w_{\rm eff}(H) \equiv \frac{p_{\rm eff}}{\rho_{\rm DE}}	\simeq -1 + \varepsilon	- \frac{3 \, \zeta_{\rm eff}(H) \, H}{T_s}
	\label{weff-general}
\end{equation}
Substituting the explicit form of $\zeta_{\rm eff}(H)$ from \eqref{effective viscosity} gives,

\begin{equation}
	w_{\rm eff}(H) \simeq  -1 + \varepsilon	- 3 \frac{K_{\rm ph} \, H \, \tau(H)}{T_s(1 + H^2 \tau(H)^2)} \simeq -1 + \varepsilon - 3 \, \kappa \, \frac{H \, \tau(H)}{1 + H^2 \tau(H)^2}
	\label{weff-Mmaxwell}
\end{equation}
where in the last step we used $K_{\rm ph} = \kappa \, T_s$. Note that the function,
\begin{equation}
	f(x) \equiv \frac{x}{1 + x^2} \, , \qquad x \equiv H \, \tau(H) 
	\label{f-x-def}
\end{equation}
satisfies $0 \leq f(x) \leq 1/2$ with a maximum at $x = 1$. The viscous correction in \eqref{weff-Mmaxwell} is therefore bounded by,

\begin{equation}
	0 \leq 3 \, \kappa \, \frac{H \, \tau(H)}{1 + H^2 \tau(H)^2}
	\leq \frac{3}{2} \, \kappa 
	\label{viscous-bound}
\end{equation}
During an expanding phase with $H > 0$ and positive $\kappa$ and $\tau(H)$ the viscous term always lowers $w_{\rm eff}(H)$ relative to the non viscous value \eqref{wDE-epsilon}. A mild phantom regime with $w_{\rm eff}(H) < -1$ is therefore obtained whenever,

\begin{equation}
	3 \, \kappa \, \frac{H \, \tau(H)}{1 + H^2 \tau(H)^2} > \varepsilon
	\label{phantom-condition}
\end{equation}
Since $f(x)$ peaks near $x = 1$, the phantom deviation of $\omega_{\rm eff}$ is most pronounced when the relaxation time satisfies
\begin{equation}
	H \, \tau(H) \sim 1 \label{Htau-unity}
\end{equation}
and it is strongly suppressed in the regimes $H \tau(H) \ll 1$ (high temperature) and $H \tau(H) \gg 1$ (low temperature).
\\ \\
We assume, as mentioned previously in Sec.~3.3, that $T_H < m_{\phi}$ when DE domination or even cosmic acceleration starts (that is $z_\star > 0.6$) so that $H \tau$ is larger than unity for all $z < z_\star$. After this, as $T_H$ keeps falling with fall in the Hubble parameter $H$, the quantity $H \tau$ keeps increasing and the effective viscosity $\zeta_{\rm eff}$ as dictated by equation \eqref{effective viscosity} falls rapidly and $w_{\rm eff}$ approximately approaches $-1 + \varepsilon$ in the infinite future. This implies that we have a dynamical equation of state for dark energy in accordance with the recent findings from DESI. We will produce a basic fit for the dark energy equation of state $w_{\rm eff}$ obtained from our phonon induced brane viscosity model (equation \eqref{weff-Mmaxwell}) to DESI results in the next Section by fixing the parameters $\kappa, \varepsilon$ and $H_\star$.

\section{\centering Compressed Likelihood fit in the $(w_0 , w_a)$ Plane}
\label{sec:w0wa_compressed_fit}

\subsection{\centering Scope and Motivation}
\label{subsec:w0wa_scope}

In this section we compare the phonon model to late time expansion history constraints that are commonly summarized in the CPL plane $(w_0 , w_a)$.
The aim is to test whether the model can reproduce the region of the $(w_0 , w_a)$ posterior preferred by current data, using a transparent and reproducible compressed likelihood.
\\ \\
We do not evaluate the BAO, supernova, or CMB likelihoods directly.
Instead, we use publicly released MCMC chains for $w_0 w_a$CDM and compress their joint posterior in $(w_0 , w_a)$ to a bivariate Gaussian. The chains used here correspond to a combined analysis of DESI DR1 BAO with Pantheon+ Type Ia supernovae and Planck 2018 CMB likelihood components. Therefore, this section fits a Gaussian compression of the published joint posterior from that dataset combination.

\subsection{\centering External Posterior Chains and Gaussian Compression in $(w_0 , w_a)$}
\label{subsec:w0wa_compression}

We use the DESI DR1 BAO cosmology value added products that provide posterior chains for several cosmological models. We select cobaya MCMC chains for the $w_0 w_a$CDM extension of flat $\Lambda$CDM. An ancillary python script \texttt{Statistical Fit.py} downloads the chain files, combines them and from the posterior samples, computes the weighted mean vector,

\begin{equation}
	\hat{\theta} \; = \; (\hat{w}_0 , \hat{w}_a)^{T}
\end{equation}
and the weighted covariance matrix $\mathbf{C}$. We retain the off diagonal covariance and do not assume that $w_0$ and $w_a$ are independent. The DESI DR1 compressed likelihood products used in this work are publicly available. The analysis code script and derived output files are accessible as a \texttt{.zip} file at (\cite{KhanMGKSPDataset}).
\\ \\
We obtain the following from running ancillary code script on cobaya MCMC chains,

\begin{equation}
	\hat{\theta}
	\;=\;
	\begin{pmatrix}
		\hat{w}_0\\
		\hat{w}_a
	\end{pmatrix}
	\;=\;
	\begin{pmatrix}
		-0.828209\\
		-0.744750
	\end{pmatrix},
	\qquad
	\mathbf{C}
	\;=\;
	\begin{pmatrix}
		0.004112 & -0.016695\\
		-0.016695 & 0.085227
	\end{pmatrix}
	\label{eq:w0wa_compressed_theta_cov}
\end{equation}

\subsection{\centering Model Prediction $w_{\rm eff}(z)$ and mapping to an Effective CPL Pair}
\label{subsec:w0wa_model_to_cpl}

Recall that the phonon model predicts an effective dark energy equation of state as a function of the Hubble rate,

\begin{equation}
	w_{\rm eff}(H) \; = \; -1 \, + \, \varepsilon \, - \, 3 \, \kappa \,
	\frac{x_{\rm ph}(H)}{1 + x_{\rm ph}^{2}(H)} \, , \qquad x_{\rm ph}(H) \; \equiv \; H \, \tau_{\rm ph}(H)
	\label{eq:weff_of_H}
\end{equation}
The ancillary script uses the same relaxation model as the main text, written in dimensionless form with $h \equiv H/H_0$ and $h_\star \equiv H_\star/H_0$ and therefore,

\begin{equation}
	x_{\rm ph}(h) \; = \; \frac{h_\star}{h} \exp \left(\frac{h_\star}{h} - 1\right)
	\label{eq:xph_of_h}
\end{equation}
To obtain $w_{\rm eff}(z)$ we use a lightweight background mapping based on a flat $\Lambda$CDM expansion rate with fixed $\Omega_{m0} = 0.3$,

\begin{equation}
	h^{2}_{\Lambda}(z) \; \equiv \; \left(\frac{H_{\Lambda}(z)}{H_0} \right)^{2} \; = \; \Omega_{m0} (1+z)^{3} + (1-\Omega_{m0}) \, , \qquad \Omega_{m0} = 0.3
	\label{eq:h_lcdm}
\end{equation}
We then evaluate $w_{\rm eff}(z)$ by composing equations \eqref{eq:weff_of_H} to \eqref{eq:h_lcdm}.
\\ \\
To compare the model to the CPL plane we fit the model prediction to the CPL form (\cite{ChevallierPolarski2001, Linder2003}),

\begin{equation}
	w_{\rm CPL}(z) \; = \; w_0 \, + \, w_a \, \frac{z}{1+z}
	\label{eq:wCPL}
\end{equation}
Define $y(z) \equiv z/(1+z)$ so that $w_{\rm CPL}(z) = w_0 + w_a \, y(z)$.
Over a dense grid $0 \le z \le 1.6$ with $N = 600$ points, the script performs an ordinary least squares regression of $w_{\rm eff}(z)$ onto the basis $\{1 , y(z)\}$ and returns the effective CPL pair,

\begin{equation}
	\theta_{\rm mod}(\varepsilon,\kappa,h_\star) \; \equiv \;
	\begin{pmatrix}
		w^{\rm mod}_0\\
		w^{\rm mod}_a
	\end{pmatrix}
	\label{eq:theta_mod_def}
\end{equation}

\subsection{\centering Compressed Gaussian Likelihood and $\chi^{2}$ Statistic}
\label{subsec:w0wa_chi2}

We approximate the posterior in the $(w_0 , w_a)$ plane by a bivariate Gaussian described by equation \eqref{eq:w0wa_compressed_theta_cov}.
For a model prediction $\theta_{\rm mod}$ we define,

\begin{equation}
	\chi^{2}(\varepsilon,\kappa,h_\star) \; = \; \Delta \theta^{\,T} \, \mathbf{C}^{-1} \, \Delta \theta \, , \qquad \Delta \theta \; = \; \theta_{\rm mod} - \hat{\theta}
	\label{eq:chi2_w0wa}
\end{equation}
This quantity is the squared Mahalanobis distance between the model implied CPL pair and the observational mean. In our analysis we scan over internal model parameters to locate points whose CPL image lies close to the DESI mean, rather than performing a full likelihood fit to the underlying BAO, supernova, and CMB data. Therefore a very small value $\chi^2 \ll 1$ should be read as a consistency and viability statement at the level of the compressed summary, not as a conventional goodness of fit with an associated $p$ value. 
\\ \\
The parameter scan in \texttt{Statistical Fit.py} imposes the basic physical prior,

\begin{equation}
	0 \; < \; \kappa \; \le \; \varepsilon
	\label{eq:basic_prior_kap_le_eps}
\end{equation}
 The condition $\varepsilon \geq \kappa$ together with $\varepsilon < 1$ and $\kappa < 1$ prevents tachyonic behavior in the effective phonon description and is reasonable because $\varepsilon$ and $\kappa$ are defined as fractional response coefficients, so values above unity would correspond to an unphysical over response in a minimal model. A definitive statistical assessment requires a self consistent solution for $H(z)$ within the model and a direct fit to the full likelihoods, which we leave to future work. The scan searches within the box,
 
\begin{equation}
	0.05 \le \varepsilon \le 0.95 \, , \qquad 0.05 \le \kappa \le 0.95 \, , \qquad 0.5 \le h_\star \le 4.0
	\label{eq:scan_box}
\end{equation}
A coarse grid scan is followed by local refinements around the best point.
This design keeps the analysis fast and fully reproducible.

\subsection{\centering Numerical Results}
\label{subsec:w0wa_results}

For the DESI DR1 compressed summary in equation\eqref{eq:w0wa_compressed_theta_cov}, the ancillary script finds the best fit,

\begin{equation}
	(\varepsilon,\kappa,h_\star)_{\rm best} \; = \; (0.596835,\; 0.574051,\; 1.794118)
	\label{eq:w0wa_best_params}
\end{equation}
which maps to,

\begin{equation}
	(w^{\rm mod}_0, w^{\rm mod}_a)_{\rm best} \; = \; (-0.829119,\; -0.744835) \, , \qquad \chi^{2}_{\rm best} \; = \; 1.02088 \times 10^{-3}
	\label{eq:w0wa_best_w0wa}
\end{equation}
It is also useful to report the implied phonon sound speed diagnostic,

\begin{equation}
	c_{s}^{2} \; \equiv \; \frac{\kappa}{\varepsilon}
	\label{eq:cs2_def}
\end{equation}
which for the best fit gives,

\begin{equation}
	c_{s}^{2} \; \simeq \; 0.961825 \, , \qquad \frac{\varepsilon - \kappa}{\varepsilon} \; \simeq \; 0.038175
	\label{eq:w0wa_best_cs2}
\end{equation}
No additional near luminal band is imposed in the scan beyond $\kappa \le \varepsilon$. The closeness of $c_s^{2}$ to unity is therefore an output of the fit rather than a built in restriction.

\begin{table}[htbp]
	\centering
	\small
	\setlength{\tabcolsep}{4pt}
	\renewcommand{\arraystretch}{1.15}
	\begin{tabular}{@{}lcccc@{}}
		\hline
		Case
		& $w_0$
		& $w_a$
		& $\Delta w_0 \equiv w_0 - \hat{w}_0$
		& $\Delta w_a \equiv w_a - \hat{w}_a$
		\\
		\hline
		DESI compressed mean $\hat{\theta}$
		& $-0.828209$
		& $-0.744750$
		& $0$
		& $0$
		\\
		Best fit
		& $-0.829119$
		& $-0.744835$
		& $-9.10\times 10^{-4}$
		& $-8.45\times 10^{-5}$
		\\
		\hline
	\end{tabular}
	\caption{Comparison between the compressed CPL mean $(\hat{w}_0,\hat{w}_a)$ and the model implied CPL pair obtained from the regression procedure in Sec.\ \ref{subsec:w0wa_model_to_cpl}.}
	\label{tab:w0wa_comparison}
\end{table}
\FloatBarrier

\begin{table}[htbp]
	\centering
	\small
	\setlength{\tabcolsep}{4pt}
	\renewcommand{\arraystretch}{1.15}
	\begin{tabular}{@{}lcccccc@{}}
		\hline
		Case
		& $\varepsilon$
		& $\kappa$
		& $h_\star$
		& $c_s^2=\kappa/\varepsilon$
		& $(\varepsilon-\kappa)/\varepsilon$
		& $\chi^2$
		\\
		\hline
		Best fit
		& $0.596835$
		& $0.574051$
		& $1.794118$
		& $0.961825$
		& $0.038175$
		& $1.02088\times 10^{-3}$
		\\
		\hline
	\end{tabular}
	\caption{Summary of the unrestricted best fit in the compressed $(w_0 , w_a)$ likelihood. The scan imposes only $0 < \kappa \le \varepsilon$.}
	\label{tab:w0wa_param_summary}
\end{table}
\FloatBarrier

\subsection{\centering Pointwise Comparison of CPL Reconstructions and the model implied CPL Pair}
\label{subsec:w0wa_pointwise_comparison}

In order to visualize how close the model implied CPL pair is to the compressed DESI mean in a more direct way, it is useful to compare the corresponding CPL equations of state at representative redshifts in the DESI leverage range. Given a CPL pair $(w_0,w_a)$ we define,

\begin{equation}
	\omega(z) \; = \; w_0 \, + \, w_a \, \frac{z}{1+z}
	\label{eq:cpl_w_of_z}
\end{equation}
We evaluate equation \eqref{eq:cpl_w_of_z} for the DESI compressed mean $(\hat{w}_0,\hat{w}_a)$ from equation \eqref{eq:w0wa_compressed_theta_cov} and for the model implied CPL pair $(w_0^{\rm mod},w_a^{\rm mod})$ from equation \eqref{eq:w0wa_best_w0wa}. Table~\ref{tab:wz_pointwise} shows $\omega(z)$ and $\omega_{\rm eff}(z)$ at a set of representative redshifts $z = 0.5$ and $z = 1.0$ and in the interval $1.1 \le z \le 1.5$ in steps of $0.1$. The final column reports the difference $\Delta \omega(z) \equiv \omega_{\rm eff}(z) - \omega(z)$. The near equality of the two columns reflects the fact that the best fit lies extremely close to the compressed mean in the $(w_0,w_a)$ plane.

\begin{figure}[htbp]
	\centering
	\includegraphics[width=0.78\linewidth]{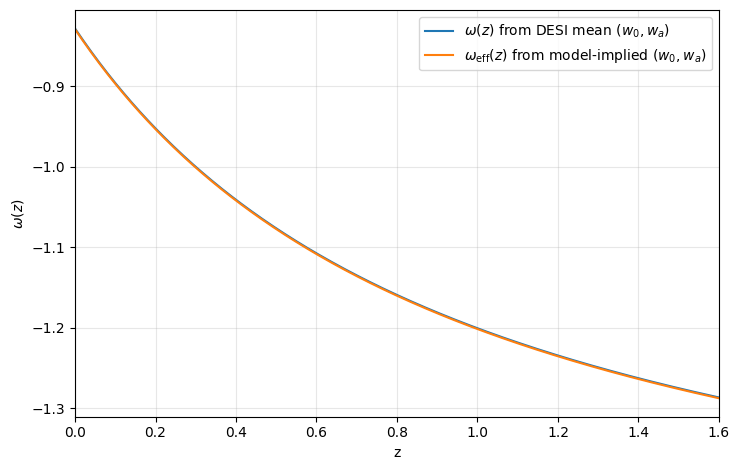}
	\caption{Overlay of the CPL equation of state $\omega(z)$ reconstructed from the DESI compressed mean $(\hat{w}_0,\hat{w}_a)$ and the model implied CPL equation of state $\omega_{\rm eff}(z)$ computed from $(w_0^{\rm mod},w_a^{\rm mod})$.}
	\label{fig:wz_cpl_overlay}
\end{figure}

\begin{table}[htbp]
	\centering
	\small
	\setlength{\tabcolsep}{4pt}
	\renewcommand{\arraystretch}{1.15}
	\begin{tabular}{@{}cccc@{}}
		\hline
		$z$
		& $\omega(z)$ from $(\hat{w}_0,\hat{w}_a)$
		& $\omega_{\rm eff}(z)$ from $(w_0^{\rm mod},w_a^{\rm mod})$
		& $\Delta \omega(z)$
		\\
		\hline
		$0.5$
		& $-1.076509$
		& $-1.077398$
		& $-8.89\times 10^{-4}$
		\\
		$1.0$
		& $-1.200584$
		& $-1.201537$
		& $-9.53\times 10^{-4}$
		\\
		$1.1$
		& $-1.218985$
		& $-1.219947$
		& $-9.62\times 10^{-4}$
		\\
		$1.2$
		& $-1.234891$
		& $-1.235855$
		& $-9.64\times 10^{-4}$
		\\
		$1.3$
		& $-1.248762$
		& $-1.249726$
		& $-9.64\times 10^{-4}$
		\\
		$1.4$
		& $-1.260957$
		& $-1.261919$
		& $-9.62\times 10^{-4}$
		\\
		$1.5$
		& $-1.271761$
		& $-1.272721$
		& $-9.60\times 10^{-4}$
		\\
		\hline
	\end{tabular}
	\caption{Pointwise comparison of CPL equations of state. The second column uses the DESI compressed mean $(\hat{w}_0,\hat{w}_a)$, while the third column uses the model implied CPL pair $(w_0^{\rm mod},w_a^{\rm mod})$. The final column shows $\Delta \omega(z) \equiv \omega_{\rm eff}(z) - \omega(z)$.}
	\label{tab:wz_pointwise}
\end{table}
\FloatBarrier
\noindent Figure \ref{fig:wz_cpl_overlay} provides the corresponding continuous comparison over $0 \le z \le 1.5$. The two curves are visually indistinguishable on the plot, which is consistent with the small values of $\Delta \omega(z)$ in Table~\ref{tab:wz_pointwise} and with the very small best fit $\chi^2$ reported in equation \eqref{eq:w0wa_best_w0wa}.

\FloatBarrier

\subsection{\centering Interpretation and Limitations of the Compressed Approach}
\label{subsec:w0wa_interpretation}

The extremely small value of $\chi^{2}_{\rm best}$ indicates that there exists a physically allowed region of $(\varepsilon,\kappa,h_\star)$ for which the model implied CPL pair is essentially coincident with the mean of the compressed posterior in the $(w_0 , w_a)$ plane. This should be interpreted as a consistency check that the phonon model can reproduce the central region of the published $w_0 w_a$ constraints for the chosen dataset combination. It is worth emphasizing that the best fit lies in a physically natural regime for the phonon interpretation. The scan is allowed to explore a broad box of parameters, subject only to the basic causality prior $0 < \kappa \le \varepsilon$, yet it selects a point with $\kappa < \varepsilon < 1$ and a small fractional splitting $\dfrac{\varepsilon - \kappa}{\varepsilon} \;\simeq\; 0.038$, which corresponds to a near luminal phonon propagation speed $c_s^2 = \kappa/\varepsilon \simeq 0.962$. This outcome is particularly appealing because near luminal propagation is precisely what one expects for very soft infrared oscillations of a stiff brane like medium. In our setup the characteristic scale $H_\star$ is of order the Hubble rate at redshift of order unity, so the associated phonon mass gap $m_\phi = H_\star/(2\pi)$ is extremely small in absolute units and only a fraction of $H_0$ (see below). The resulting Compton wavelength is therefore of order the Hubble horizon, meaning that the excitations relevant for the transient viscous response are horizon scale collective modes rather than microscopic degrees of freedom as mentioned earlier. For such long wavelength modes, relativistic propagation is the generic expectation unless the medium contains significant additional structure that enforces a much lower sound speed. In this sense the compressed likelihood fit does not merely accommodate near luminal phonons, it points to them. The parameter scan naturally identifies a region in which the brane is strongly elastic but still well behaved, with $\varepsilon$ and $\kappa$ both below unity and separated by only a few percent, so that the phonons remain causal while being very close to the luminal limit.
\\ \\ 
At the same time, several limitations should be kept in mind. First, the Gaussian compression uses only the mean and covariance of $(w_0 , w_a)$, so it discards non Gaussian structure and correlations with other cosmological parameters present in the full chains. Second, the mapping from the model to $(w_0 , w_a)$ depends on the chosen regression range $0 \le z \le 1.6$ and on the use of an unweighted least squares fit to $w_{\rm eff}(z)$. Third, we evaluate $w_{\rm eff}(z)$ on a fixed $\Lambda$CDM background with $\Omega_{m0}=0.3$ rather than solving a fully self consistent background evolution for the model. These choices are deliberate in order to keep the test lightweight and reproducible, and they are appropriate for an initial comparison in the CPL plane.

\subsection{\centering Asymptotic Equation of State of Dark Energy}
Another important point concerns the asymptotic behaviour of the effective dark energy sector in the infinite past and in the infinite future in the context of the phonon model. A useful diagnostic of the far past and far future limits is provided by the dimensionless relaxation variable,

\begin{equation}
	x_{\rm ph}(H) \equiv H \, \tau_{\rm ph}(H) = \frac{H_\star}{H}\exp \left(\frac{H_\star}{H} - 1\right)
\end{equation}
In the asymptotic past, $H \gg H_\star$, we have $H_\star/H \to 0$ and therefore $x_{\rm ph}(H) \to 0$. In the asymptotic future, $H \ll H_\star$ so $x_{\rm ph}(H) \to 0$. In both limits the Maxwell kernel appearing in equation (\ref{eq:weff_of_H}) satisfies,

\begin{equation}
	\frac{x_{\rm ph}(H)}{1 + x_{\rm ph}^{2}(H)} \; \longrightarrow \; 0
	\qquad \text{as} \qquad H \to \infty \quad \text{or} \quad H \to 0
\end{equation}
which implies that the relaxation sector decouples in the far past and far future. Using equation (3.42), whose dependence on the relaxation dynamics enters only through $x_{\rm ph}(H)$, we therefore obtain the same asymptotic value of the effective equation of state at both ends of the cosmological evolution,

\begin{equation}
	\lim_{H \to \infty} w_{\rm eff} = \lim_{H \to 0} w_{\rm eff} = -1 + \varepsilon
\end{equation}
In our least-squares fit to the DESI-motivated CPL targets we find $\varepsilon \simeq 0.597$, and hence an asymptotic dark energy equation of state,

\begin{equation}
	w_{\rm eff} \; \to \; -1 + \varepsilon \; \simeq \; -0.403 \; \simeq \; -0.40
\end{equation}
both in the early universe and in the infinite future. The bulk viscous contribution in equation (\ref{eq:weff_of_H}) is therefore a genuinely transient effect that operates only while $x_{\rm ph}(H)$ is of order unity, that is, while the Hubble rate is comparable to the mass gap scale $H_\star$. In this intermediate regime the effective equation of state can dip below $w=-1$ and track the DESI-driven CPL reconstruction, while at very early and very late times the model relaxes back to a regular quintessence-like value $w_{\rm eff} \simeq -0.40$.

\subsection{\centering Reproducibility and Ancillary Material}
\label{subsec:w0wa_reproducibility}

As mentioned earlier, all numbers reported in this section are produced by the ancillary script \texttt{Statistical Fit.py}. The script downloads the public DESI DR1 VAC chain files for $w_0 w_a$CDM for the dataset combination encoded in the base URL, computes the weighted mean and covariance of $(w_0 , w_a)$, performs the mapping,

\begin{equation}
	(\varepsilon,\kappa,h_\star) \; \longrightarrow \; w_{\rm eff}(z) \; \longrightarrow \; (w_0^{\rm mod}, w_a^{\rm mod}),
\end{equation}
and evaluates and minimizes $\chi^{2}$ in equation \eqref{eq:chi2_w0wa}.
For exact numerical reproducibility, the script writes the file \texttt{w0wa\_compressed.tex} and a JSON summary file \texttt{fit\_results.json}. The ancillary code file and its outputs are available as a \texttt{.zip} file at (\cite{KhanMGKSPDataset}).

\subsection{\centering Phonon Mass Scale associated with $H_\star$}
\label{subsec:w0wa_mass_scale}

In the simplified description of the relaxation time, the characteristic scale $H_\star$ can be interpreted as a phonon mass gap via,

\begin{equation}
	m_{\phi} \; = \; \frac{H_\star}{2\pi}
	\label{eq:mphi_def}
\end{equation}
Using the best fit value $h_\star = H_\star/H_0 = 1.794118$, this gives,

\begin{equation}
	m_{\phi} \; = \; \frac{h_\star}{2\pi} \, H_0 \; \simeq \; 0.28554 \, H_0
\end{equation}
Adopting $H_0 \simeq 1.4 \times 10^{-33}\,\mathrm{eV}$, we obtain
\begin{equation}
	m_{\phi} \; \simeq \; 4.00 \times 10^{-34} \,\mathrm{eV} \; \simeq \;
	7.13 \times 10^{-70} \,\mathrm{kg}
\end{equation}
This mass scale is around 30\% of the present Hubble scale, so the associated Compton wavelength is of order the cosmological horizon.
This is consistent with the interpretation of the phonons as infrared degrees of freedom that respond to the global expansion and become most relevant when $H$ is comparable to the mass gap scale.

\section{\centering Conclusion and Outlook}
In this work we modeled cosmic acceleration as the effective dynamics of an elastic brane that represents space itself. The underlying geometric contribution is described by a Nambu--Goto type tension $T_s$ that survives as a residual cosmological constant in a minimal sequester framework. On top of this background tension we introduced a phonon fluid built from three scalar fields $\phi^I$ and an invariant $b = \sqrt{\det B_{IJ}}$ that controls the effective action $F(b)$. At the background level this construction reproduces a perfect fluid with energy density $\rho_{\rm ph}(b)$, pressure $p_{\rm ph}(b)$ and bulk modulus $K_{\rm ph}$. We parametrized the phonon sector by two dimensionless constants $\epsilon$ and $\kappa$ which fix the ratio $(\rho_{\rm ph}+p_{\rm ph})/T_s$ and the bulk modulus in units of the space tension. This led to a simple and transparent relation for the phonon sound speed, $c_s^2 = \kappa/\epsilon$, with stability requiring $0 < \kappa/\epsilon \leq 1$.
\\ \\
We then studied dissipative corrections that arise from bulk viscosity in the phonon fluid. Using the Kubo picture as motivation we treated the response of the isotropic stress to a metric perturbation within a Maxwell model, in which the viscous pressure relaxes on a time scale $\tau(H)$ that depends on the Hubble rate. The effective bulk viscosity $\zeta_{\rm eff}(H)$ is suppressed at high frequency and is maximal when the product $H\tau(H)$ is of order unity. For the purpose of phenomenology we adopted a simple ansatz for the relaxation time that follows from a Boltzmann suppressed scattering rate at a characteristic scale $H_\star$. This leads to a compact expression for the effective dark energy equation of state,
\begin{equation}
	w_{\rm eff}(H) = -1 + \epsilon - 3\kappa \, \frac{x(H)}{1 + x(H)^2} \, , \quad x(H) = H \, \tau(H)
\end{equation}
which makes it clear that a transient phantom phase can occur when $x(H)$ passes through unity, while $w_{\rm eff}$ tends to $-1+\epsilon$ at very early and very late times.
\\ \\
Using a compressed Gaussian likelihood in the CPL plane derived from publicly released posterior chains for DESI DR1 BAO combined with Pantheon+ and Planck 2018, we find that the model can reproduce the central region of the published $(w_0,w_a)$ constraints without imposing any additional near luminal band beyond the basic causality prior $0 < \kappa \le \varepsilon$. In particular, the parameter scan naturally selects a best fit with $\kappa < \varepsilon < 1$ and a small fractional splitting $(\varepsilon-\kappa)/\varepsilon \simeq 0.038$, which implies a near luminal sound speed $c_s^2 = \kappa/\varepsilon \simeq 0.962$ as an output of the fit. This behaviour is physically natural in the present framework because the relevant phonon excitations are infrared collective oscillations of the brane with a mass gap $m_\phi = H_\star/(2\pi)$ set by a Hubble scale of order unity in redshift, so their Compton wavelength is of order the cosmological horizon. For such horizon scale modes, relativistic propagation is the generic expectation unless additional microphysical structure enforces a significantly smaller sound speed. In this sense, the compressed likelihood comparison does not merely accommodate near luminal phonons, it points to them in a particularly economical way.
\\ \\
Several natural directions for future work emerge from this first exploration. The most important step is to derive the bulk viscosity $\zeta(H)$ and the relaxation time $\tau(H)$ from a more complete microscopic theory of the phonon sector. One would like to start from an explicit effective field theory for the longitudinal modes of the brane or for additional scalar degrees of freedom, compute their dispersion relations and interaction rates, and evaluate the stress tensor correlator that enters the Kubo formula. This program would allow one to determine $\zeta(H)$ and $\tau(H)$ from first principles rather than from phenomenology and would test whether the simple scaling used here is robust or only a useful approximation. A more complete theory could clarify how the NG brane tension and the phonon fluid arise from a fundamental microscopic description and whether similar viscoelastic effects appear in other approaches to late time acceleration. The results presented here indicate that an elastic and viscous space is a viable candidate for the dark energy sector and they motivate a more systematic study of the microphysics and cosmological implications of such models.

\section{\centering Acknowledgment}
The author acknowledges the use of \textit{ChatGPT-5.1} (November 2025 version) to edit and refine the tone and clarity of selected sections of this manuscript. The model was used solely to polish and rephrase text originally written by the author. All passages prepared with LLM assistance were reviewed, revised where needed and approved by the author, who retains full responsibility for the final content.

\newpage 

\section*{\centering Appendix A}

The covariant divergence of \(J^\mu\) is,

\begin{equation}
	\nabla_\mu J^\mu = \frac{1}{6} \nabla_\mu
	\Bigl( \epsilon^{\mu\nu\rho\sigma} \epsilon_{IJK}
	\,\partial_\nu \phi^I \partial_\rho \phi^J \partial_\sigma \phi^K \Bigr)
\end{equation}
The Levi Civita tensor density is covariantly constant,
\(\nabla_\mu \epsilon^{\mu\nu\rho\sigma} = 0\) and \(\phi^I\) are scalar fields, so \(\nabla_\mu \partial_\nu \phi^I = \nabla_\nu \partial_\mu \phi^I\). Using these facts we can write,

\begin{align}
	\nabla_\mu J^\mu = \frac{1}{6} \epsilon^{\mu\nu\rho\sigma} \epsilon_{IJK} \Bigl[ & (\nabla_\mu \partial_\nu \phi^I) \partial_\rho \phi^J \partial_\sigma \phi^K + \partial_\nu \phi^I (\nabla_\mu \partial_\rho \phi^J) \partial_\sigma \phi^K \notag \\ 
	& + \partial_\nu \phi^I \partial_\rho \phi^J (\nabla_\mu \partial_\sigma \phi^K) \Bigr]
\end{align}
Each term in brackets contains a symmetric pair of derivative indices
inside an antisymmetric contraction with \(\epsilon^{\mu\nu\rho\sigma}\). For example in the first term, we have
\(\nabla_\mu \partial_\nu \phi^I = \nabla_\nu \partial_\mu \phi^I\). So if we exchange \(\mu\) and \(\nu\) in that term, the Levi Civita tensor picks up a minus sign but the derivative piece stays the same, so the term cancels itself. The same reasoning applies to the second term
which is symmetric in \(\mu\) and \(\rho\) and to the third term which is symmetric in \(\mu\) and \(\sigma\). All three contributions cancel and one finds,

\begin{equation}
	\nabla_\mu J^\mu = 0
\end{equation}
This conservation law expresses the fact that the number of comoving volume elements of the medium is conserved.

\section*{\centering Appendix B}
The quantity \(J_\mu J^\mu\) is a scalar and the term \(b^2 = \det B^{IJ}\) is also a scalar. If two scalar expressions agree in one frame at a point then they agree in every frame at that point. We can therefore prove the identity in a convenient local frame.
\\ \\
We choose a local inertial frame where the metric at a given spacetime point is,

\begin{equation}
	g_{\mu\nu} = \mathrm{diag}(-1, 1, 1, 1)
\end{equation}
and choose comoving coordinates for the background configuration of the medium, that is,

\begin{equation}
	\phi^I = x^I
\end{equation}
with \(I = 1,2,3\). In this frame,

\begin{equation}
	\partial_0 \phi^I = 0, \qquad \partial_i \phi^J = \delta_i^{\ J}
\end{equation}
where Latin indices \(i,j,k\) run over spatial components. The matrix \(B^{IJ}\) is then,

\begin{equation}
	B^{IJ} = g^{\mu\nu} \partial_\mu \phi^I \partial_\nu \phi^J = g^{ij} \delta_i^{\ I} \delta_j^{\ J} = \delta^{IJ}
	\label{B ij}
\end{equation}
so,
\begin{equation}
	\det B^{IJ} = 1, \qquad	b = 1
\end{equation}
Next we evaluate \(J^\mu\). For the time component,

\begin{equation}
	J^0	= \frac{1}{6} \epsilon^{0ijk} \epsilon_{IJK} \, \partial_i \phi^I \partial_j \phi^J \partial_k \phi^K	= \frac{1}{6} \epsilon^{0ijk} \epsilon_{ijk}
\end{equation}
where in the last step we used \(\partial_i \phi^J = \delta_i^{\ J}\). There are \(3!\) non zero terms in the sum over the Levi-Cevita tensor so,

\begin{equation}
	J^0 = 1
\end{equation}
For the spatial components we have,

\begin{equation}
	J^i	= \frac{1}{6} \epsilon^{i\nu\rho\sigma} \epsilon_{IJK}
	\partial_\nu \phi^I \partial_\rho \phi^J \partial_\sigma \phi^K
\end{equation}
In our configuration one of the derivatives is always \(\partial_0 \phi^I = 0\), so every term vanishes and we obtain,

\begin{equation}
	J^i = 0
\end{equation}
Thus,

\begin{equation}
	J^\mu = (1, 0, 0, 0)
\end{equation}
The norm is,

\begin{equation}
	J_\mu J^\mu = g_{00} (J^0)^2 = -1
\end{equation}
Since \(b = 1\) in this frame, we have,

\begin{equation}
	J_\mu J^\mu = - b^2
\end{equation}
at this point and because both sides are scalars this equality holds in any frame and in any coordinate system.
\\ \\
We can therefore also define the fluid four velocity as
\begin{equation}
	u^\mu \equiv \frac{J^\mu}{b} \, , \quad u^\mu = (1, 0, 0, 0) \, , \quad u_\mu u^\mu = -1
	\label{four velocity relation}
\end{equation}

\section*{\centering Appendix C}
We now consider the tensor,

\begin{equation}
	h_{\mu\nu} \equiv (B^{-1})_{IJ} \partial_\mu \phi^I \partial_\nu \phi^J
	\label{projector result}
\end{equation}
We want to show that in fact,

\begin{equation}
	h_{\mu\nu} = g_{\mu\nu} + u_\mu u_\nu
\end{equation}
Since both sides are rank two tensors, if they agree in one frame at one point then they agree in every frame at that point. Again we work in the local inertial rest frame described above where \(g_{\mu\nu} = \mathrm{diag}(-1,1,1,1)\) and \(\phi^I = x^I\) at the point of interest. We already found from \eqref{B ij} and \eqref{four velocity relation} that,

\begin{equation}
	B^{IJ} = \delta^{IJ}, \qquad (B^{-1})_{IJ} = \delta_{IJ}, \qquad
	u^\mu = (1,0,0,0), \qquad u_\mu = (-1,0,0,0)
\end{equation}
This gives us the components of \(h_{\mu\nu}\) as,

\begin{equation}
	h_{00}= (B^{-1})_{IJ} \, \partial_0 \phi^I \partial_0 \phi^J	= \delta_{IJ} \cdot 0 \cdot 0	= 0
\end{equation}

\begin{equation}
	h_{0i} = h_{i0} = (B^{-1})_{IJ} \, \partial_0 \phi^I \partial_i \phi^J	= \delta_{IJ} \cdot 0 \cdot \delta_i^{\ J} = 0
\end{equation}

\begin{equation}
	h_{ij} = (B^{-1})_{IJ} \, \partial_i \phi^I \partial_j \phi^J 	= \delta_{IJ} \delta_i^{\ I} \delta_j^{\ J} = \delta_{ij}
\end{equation}
In the rest from for the right hand side in equation \eqref{projector result}, we have,

\begin{equation}
	u_\mu = (-1,0,0,0),	\qquad	g_{\mu\nu} = \mathrm{diag}(-1,1,1,1)
\end{equation}
Thus,

\begin{align}
	g_{00} + u_0 u_0 & = -1 + 1 = 0,	\\
	g_{0i} + u_0 u_i & = 0 + 0 = 0,	\\
	g_{ij} + u_i u_j & = \delta_{ij} + 0 = \delta_{ij}.
\end{align}
These are exactly the components of \(h_{\mu\nu}\) that we computed above. Therefore at the point under consideration, we have,

\begin{equation}
	h_{\mu\nu} = g_{\mu\nu} + u_\mu u_\nu
\end{equation}
as claimed. Since both sides transform covariantly as tensors this equality holds in any coordinate system. This shows that,

\begin{equation}
	h_{\mu\nu} \equiv (B^{-1})_{IJ} \partial_\mu \phi^I \partial_\nu \phi^J
\end{equation}
is the projector onto spatial directions orthogonal to the fluid four velocity \(u^\mu\). It satisfies,

\begin{equation}
	h_{\mu\nu} u^\nu = 0, \qquad h^\mu_{\ \alpha} h^\alpha_{\ \nu} = h^\mu_{\ \nu}
\end{equation}
as required for a projector.

\end{document}